\documentclass{mem}
\usepackage{natbib}\usepackage{txfonts}\usepackage{balance}
\usepackage{graphicx}
\usepackage[a4paper,breaklinks,dvipdfm]{hyperref}
\idline{75}{282}
\begin{document}
\def\teff{$T\rm_{eff }$}
\def\kms{$\mathrm {km s}^{-1}$}

\title{
Accretion and Outflow Activity in Brown Dwarfs
}

   \subtitle{}

\author{
B. Riaz\inst{1} 
          }

  \offprints{B. Riaz}

\institute{
Centre for Astrophysics Research, Science \& Technology Research Institute, University of Hertfordshire, Hatfield, AL10 9AB, UK.
\email{b.riaz@herts.ac.uk}
}

\authorrunning{Riaz }

\titlerunning{Accretion and Outflow Activity in BDs}

\abstract{
An investigation of the magnetospheric accretion and outflow signatures in sub-stellar objects is a natural extension of similar studies conducted on classical T Tauri stars (CTTS), and helps understand if brown dwarfs (BDs) follow the same formation mechanism as stars. Over the past decade, evidence for accretion in very low-mass stars (VLMs) and BDs has been accumulated using various techniques, which indicates that the overall accretion characteristics are continuous across the sub-stellar boundary. Outflow activity in VLMs and BDs has been confirmed based on spectro-astrometry of forbidden emission lines observed in the optical, and in millimetre continuum images of CO J=2-1 emission. This review summarizes the past and current state of observational work on accretion and outflow activity in VLMs and BDs, particularly with the advent of new instruments such as VLT/X-Shooter which has allowed the study of several accretion and outflow indicators over a wider wavelength range. 

\keywords{stars: low-mass, brown dwarfs -- accretion, accretion disks -- ISM: jets and outflows}

}

\maketitle{}

\section{Introduction}

The collapse of a cold molecular cloud core over a timescale of $\sim$10$^{5}$ yrs leads to the formation of a protostar surrounded by an infalling envelope and an accretion disk (e.g., Hartmann 2005). A now familiar picture of magnetospheric accretion in classical T Tauri stars (CTTSs) is based on the model originally proposed for neutron stars by Ghosh \& Lamb (1979), and first applied to CTTSs by Uchida \& Shibata (1985) and Bertout et al. (1988). In this model, the inner disk region is the main site of different phenomena, each of which make a contribution at different wavelengths. Magnetic field strengths of a few kG have been measured on low mass stars, which are twice as strong as the fields measured on G and K stars (e.g., Johns-Krull \& Valenti 1996). Such strong magnetic fields are capable of disrupting the accretion disk at small inner radii ($\sim$0.1 AU), causing accreting material to fall at free-fall velocities onto the star along the accretion columns. The hot ($\sim$10$^{4}$ K) gas in these accretion columns emits strongly in the Balmer and other accretion-associated lines (e.g., Muzzerolle et al. 1998; 2003). The hot continuum emission seen in the optical and the ultraviolet wavelengths is produced by the accretion energy dissipated when the hot gas shocks at the stellar surface. The accretion disk itself emits in the mid- and far-infrared from dust at a range of temperatures. This model can also explain the generation of accretion-powered outflows and winds launched from near the poles of the star (e.g., Hirose et al. 1997), and observed as forbidden emission lines. The signposts of these processes have been observed in several studies conducted on CTTSs (e.g., Hartmann et al. 1994; Gullbring et al. 1998; Hartigan et al. 1995; Meyer et al. 1997; Valenti et al. 1993; Calvet \& Gullbring 1998; Muzerolle et al. 2001), which led to the investigation of accretion activity below the sub-stellar mass limit. 

\section{Magnetospheric Accretion in Brown Dwarfs}

\begin{figure}
\resizebox{\hsize}{!}{\includegraphics[clip=true]{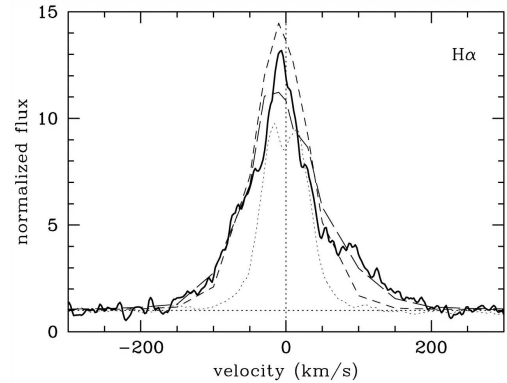}}
\caption{
\footnotesize The H$\alpha$ profile for V410 Anon 13 (solid line). The profile for a non-accreting VLMS MHO 8 is shown with dotted line. Model fits to the observed profile are shown for a disk inclination angle of 50$\degr$ (long-dashed line) and 65$\degr$ (short-dashed line). Source: Muzerolle et al. (2000).
}
\label{v410anon13}
\end{figure}

The first confirmed detection of accretion in a brown dwarf (BD) was presented by Muzerolle et al. (2000) for the Taurus member V410 Anon 13 ($\sim$0.04--0.06$M_{\sun}$; Brice\~{n}o et al. 1998). The H$\alpha$ emission line profile of V410 Anon 13 is broad and asymmetric, with most emission observed blueward of the line center (Fig.~\ref{v410anon13}). The profile width at 10\% of the peak is $\sim$250 km s$^{-1}$, much larger than the $\sim$100 km s$^{-1}$ widths observed for non-accreting and weak-line T Tauri stars (WTTSs) with narrow, symmetric H$\alpha$ profiles. The line width and asymmetries were consistent with that observed for CTTSs (e.g., Muzerolle et al. 1998), and indicated free-falling accretion flows. However, the mass accretion rate obtained from H$\alpha$ profile modeling is a small value of $\sim$5 $\times$ 10$^{-12}$ $M_{\sun}$ yr$^{-1}$, which is over 3 orders of magnitude smaller than the typical rates of $\sim$10$^{-8}$ $M_{\sun}$ yr$^{-1}$ derived for solar mass CTTSs (e.g., Gullbring et al. 1998). Also notable was the lack of any measurable veiling of the photospheric absorption lines in the V410 Anon 13 spectrum, which was reasoned to be due to the extremely small accretion rate.

This first discovery was followed by larger surveys of accretion properties in very low-mass stars (VLMs) and BDs in several star-forming regions and young associations (e.g., Muzerolle et al. 2003; 2005; White \& Basri 2003; Barrado \& Mart\'{i}n 2003; Jayawardhana et al. 2003; Natta et al. 2004; Mohanty et al. 2005). A variety in the H$\alpha$ line profiles was observed, from broad, asymmetric profiles with some exhibiting red- or blue-shifted absorption components, to narrow, symmetric line shapes. The mass accretion rates derived from the H$\alpha$ emission line were in the range of 10$^{-9.3}$--10$^{-12}$ $M_{\sun}$ yr$^{-1}$. Most of the objects with spectral type later than M4 showed insignificant continuum veiling, with the excess to photospheric flux ratios of $<$0.2 at optical wavelengths of $\sim$6000-6500 $\AA$ (e.g., Muzerolle et al. 2003; White \& Basri 2003). Based on accretion shock region models, Muzerolle et al. (2003) estimated that the surface coverage area for accretion shocks is $\leq$10$^{-3}$ smaller in sub-stellar objects compared to the typical CTTSs values, and measurable veiling is produced only when $\dot{M}_{acc}$ $>$ 10$^{-10}$ $M_{\sun}$ yr$^{-1}$. 

\subsection{Accretor/non-accretor boundary}

\begin{figure}
\resizebox{\hsize}{!}{\includegraphics{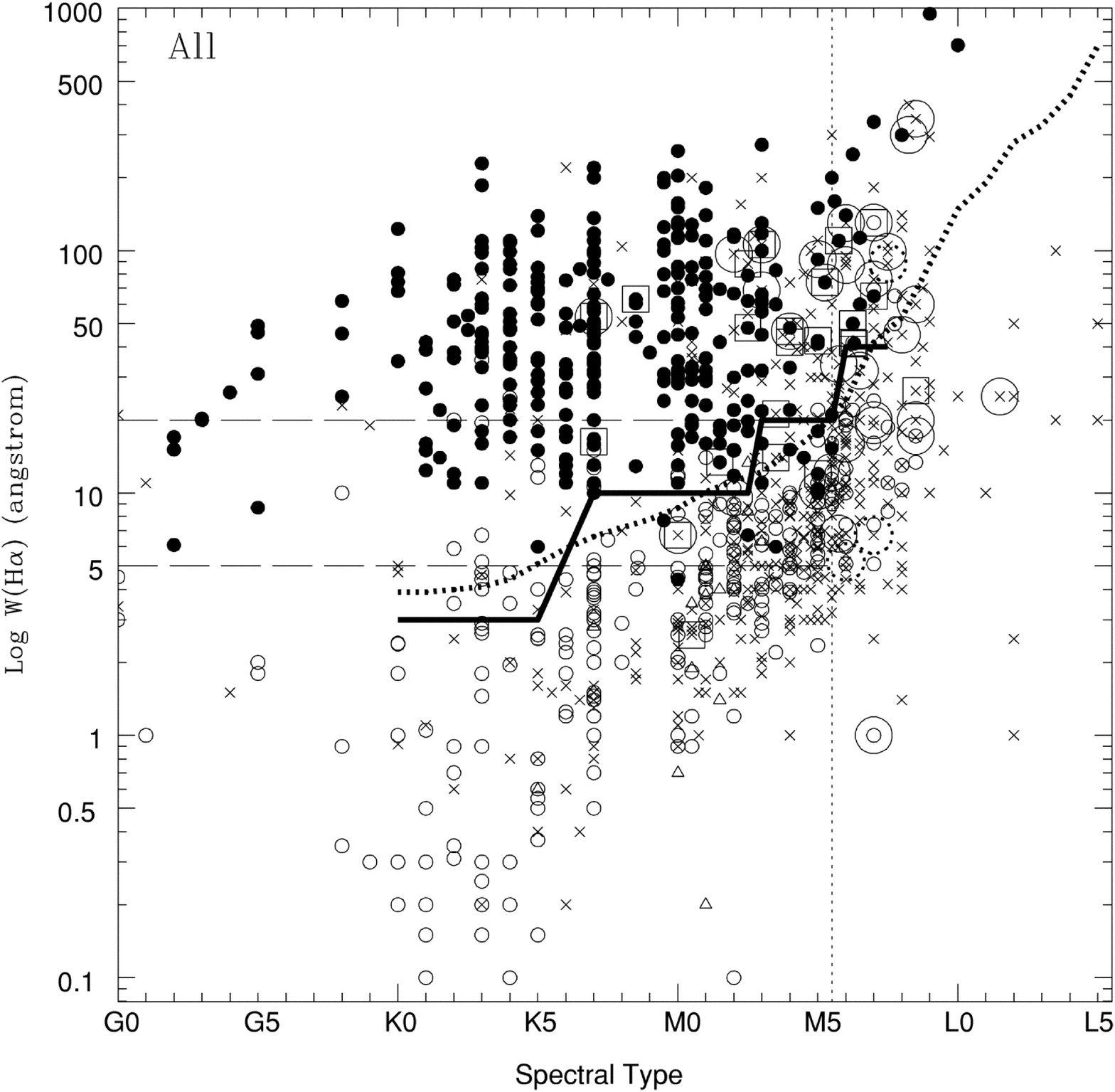}}
\caption{
\footnotesize The H$\alpha$ EW plotted against spectral type for several star-forming regions, from the work of Barrado \& Mart\'{i}n (2003). Filled and open circles represent CTTSs and WTTSs, respectively. Large open circles represent objects with mid-IR excesses. The dotted thick curve is the H$\alpha$ saturation criterion, whereas the solid line is the accretor/non-accretor threshold defined by White \& Basri (2003) based on the H$\alpha$ 10\% width. 
}
\label{bm03}
\end{figure}

Due to the very low veiling levels, the H$\alpha$ emission line equivalent width (EW) and/or the width of the H$\alpha$ line profile at 10\% of the peak intensity has been used to distinguish between accreting and non-accreting objects in VLMSs and BDs. Barrado \& Mart\'{i}n (2003) devised a threshold based on the EW of the H$\alpha$ line. They determined the average saturation limit for the H$\alpha$ luminosity in young open clusters to be log ($L_{H\alpha}$/$L_{bol}$) = -3.3, which is then converted into a quantitative EW cutoff for different spectral types (Fig.~\ref{bm03}). The limiting EW increases towards later spectral types due to the contrast effect, i.e., the same H$\alpha$ flux becomes more prominent against a cooler photosphere. Objects that lie above the cutoff EW at a given spectral type are classified as accretors, while the ones below it are non-accretors or chromospherically active objects. White \& Basri (2003) pointed out that the veiling of the photospheric absorption lines due to the excess continuum emission results in a full width of the H$\alpha$ line at 10\% of the line peak at values $\geq$ 270 km s$^{-1}$ in CTTSs. Their accretor/non-accretor threshold was therefore set at an H$\alpha$ 10\% width of $\geq$ 270 km s$^{-1}$, regardless of the spectral type. This criteria was revised for VLMSs and BDs by Muzerolle et al. (2003) and Jayawardhana et al. (2003), who noted that asymmetries in the H$\alpha$ line profile, as well as strong emission in the accretion-associated permitted lines of He I, O I, and Ca II, are observed for H$\alpha$ 10\% widths as low as 200 km s$^{-1}$. They therefore proposed a lower threshold of H$\alpha$ 10\% width of $\geq$ 200 km s$^{-1}$, which is now widely used as an accretion diagnostic for VLMSs and BDs. 

\subsection{Other accretion indicators}

\begin{figure}
\resizebox{\hsize}{!}{\includegraphics{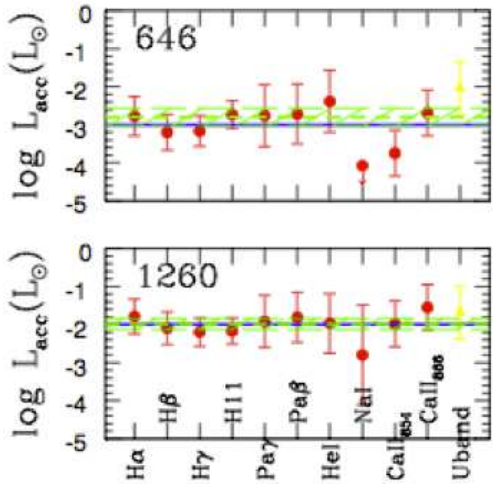}}
\caption{
\footnotesize A comparison of the accretion luminosity measured for VLMSs and BDs using 10 accretion line indicators; the average of the ten measurements is indicated by the dashed line. The point for `U-band' represents the accretion luminosity measured from the UV excess emission. Source: Rigliaco et al. (2012).
}
\label{rig}
\end{figure}

In addition to the H$\alpha$ emission line, other spectral lines have also been identified as good accretion indicators in BD spectra. Muzerolle et al. (2003; 2005) noted emission in the H$\beta$, H$\gamma$, He I $\lambda$5876, and the Ca II infrared triplet lines in nearly half of the BD accretors in their sample. Natta et al. (2004; 2006) reported emission in the hydrogen recombination line of Pa$\beta$ in 70\% of the accreting BDs in the $\rho$-Ophiuchus region, while 20\% of these accretors also showed emission in the Br$\gamma$ line. They found a correlation between the Pa$\beta$ line luminosity and the accretion luminosity derived from the H$\alpha$ profile and/or veiling measurements, similar to that observed earlier for CTTSs (e.g., Calvet et al. 2004). Mohanty et al. (2005) noted that the Ca II $\lambda$8662 line flux in VLMS and BD accretors declines sharply with the mass accretion rate derived from veiling and H$\alpha$ modeling, and the correlation is observed over $\sim$4 orders of magnitude in the line flux and accretion rate, down to an accretion rate of 10$^{-11}$ $M_{\odot}$ yr$^{-1}$. In a recent study of accretion activity in VLMSs and BDs in the $\sigma$ Orionis region using X-Shooter spectroscopy, Rigliaco et al. (2012) have identified 10 accretion line indicators, and have computed their correlations with the accretion luminosity derived from the continuum excess emission in the UV and visual (Fig.~\ref{rig}). The best correlations are observed for the hydrogen recombination lines, and these are recommended to provide good and consistent measurements of accretion luminosity for VLMSs and BDs. The correlations for the Na I $\lambda$5893 and the Ca II infrared triplet lines show larger dispersions in comparison, and it is reasoned to be due to possible contribution from the outflow emission in these lines. The accretion luminosity derived from the various line luminosities can have a spread of more than an order of magnitude, depending on the diagnostic used. 

\subsection{Accretion activity versus mass}

\begin{figure}
\resizebox{\hsize}{!}{\includegraphics{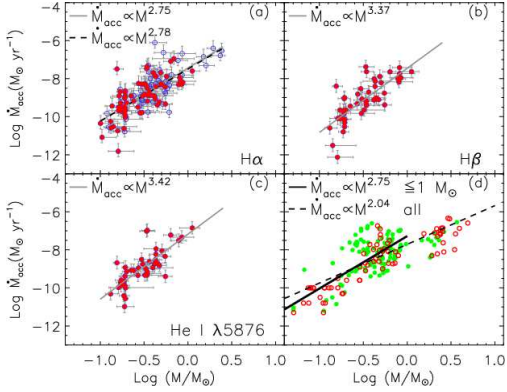}}
\caption{
\footnotesize The relation between mass accretion rate and stellar mass using different accretion indicators. Panel {\bf d} at bottom, right, shows the relation for all indicators combined; filled circles represent accretion rates measured from the $U$-band excess emission. Source: Fang et al. (2011). 
}
\label{fang}
\end{figure}

One of the main results that has been obtained from these accretion studies is the dependence of the mass accretion on the stellar mass. By more than tripling the number of known VLMS/BD accretors, Muzerolle et al. (2003; 2005) extended the trend between mass accretion rate and stellar mass earlier noted for CTTSs down to sub-stellar masses of $\sim$0.02 $M_{\odot}$. The mass accretion rates are about 2-3 orders smaller for BD accretors compared to the typical rates for CTTSs. For most BDs, the accretion rates are $<$10$^{-10}$ $M_{\odot}$ yr$^{-1}$, with the lowest values measured of $\sim$4-5$\times$10$^{-12}$ $M_{\odot}$ yr$^{-1}$ (e.g., Muzerolle et al. 2003). By combining the VLMS/BD measurements with those for CTTSs, Muzerolle et al. (2003; 2005) and Mohanty et al. (2005) derived a relation of $\dot{M}_{acc} \propto M_{*}^{2}$, which appears fairly continuous across $\sim$6 orders of magnitude in accretion rate and more than $\sim$2 orders of magnitude in mass. More recently, Fang et al. (2011) have compiled from the literature the mass accretion rates and stellar masses for young stellar objects in several star-forming regions and stellar associations, and have recomputed the relations for different accretion indicators (Fig.~\ref{fang}). For the $\dot{M}_{acc}$ vs. $M_{*}^{\alpha}$ relation, they find a range in values for the $\alpha$ exponent between 2.75 and 3.42. Within the uncertainties, the relations are similar from the three accretion line diagnostics investigated, though the relations are steeper for the H$\beta$ and the He I $\lambda$5876 lines. Combing all data together, there is a clear trend of decreasing accretion rate with decreasing mass, as noted in earlier works. There appears, however, a break in the fits around 1 $M_{\odot}$, with a steeper slope for lower mass objects. There is also a considerable spread of $\pm$ 1 dex about the mean relation in the accretion rates, which is mainly due to the different methodology used to derive the accretion rate, as these would trace different formation regions. The accretion rates determined from H$\alpha$ line modeling or other accretion line measurements would trace the infalling gas, whereas those derived from UV excess measurements would trace the accretion shock emission, and could result in comparatively higher accretion rates at a given stellar mass (e.g., filled points in Fig.~\ref{fang}). Other factors such as intrinsic variability or age differences could also play a role in the observed scatter (e.g., Muzerolle et al. 2003). 

\subsection{Accretion activity versus age}

\begin{figure}
\resizebox{\hsize}{!}{\includegraphics{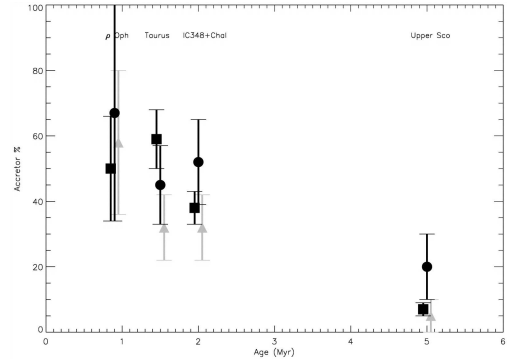}}
\caption{
\footnotesize
The disk mass accretion rate as a function of age. Filled triangles represent low-mass objects with spectral type $\geq$M5, classified as accretors using the H$\alpha$ 10\% width criteria. Filled circles represent the same set of objects classified using the H$\alpha$ EW accretor criteria. Filled squares are higher mass stars with spectral types between K0 and M4. Source: Mohanty et al. (2005).
}
\label{age}
\end{figure}

The mass accretion rate also shows a clear trend with the stellar age, as can be seen from the compiled data from various surveys presented in Mohanty et al. (2005) and Fang et al. (2011). Figure~\ref{age} shows a general trend of decreasing accretor frequency with increasing age, but the trend is more prominent for stars below $\sim$0.5$M_{\odot}$ than for more massive stars. Among stars with masses below $\sim$0.3$M_{\odot}$, the accretion rate evolves from $\sim$10$^{-8}$ -- 10$^{-9}$ $M_{\odot}$ yr$^{-1}$ for ages below $\sim$1 Myr, to $\sim$10$^{-9}$ -- 10$^{-11}$ $M_{\odot}$ yr$^{-1}$ for ages above $\sim$3Myr (e.g., Fang et al. 2011). There is a notable drop in the accretor frequency between ages of $\sim$3 and 5 Myr (Fig.~\ref{age}), suggesting that accretion activity becomes insignificant by an age of $\sim$5 Myr. It is more likely that the accretion activity drops below the measurable levels ($\sim$10$^{-12}$ $M_{\odot}$ yr$^{-1}$) by this age, and weaker accretors cannot be identified. Figure~\ref{age} also shows that while the method of classifying accretors (H$\alpha$ EW vs. 10\% width criteria) could result in a different accretor frequency at a given age, the overall trend of declining frequency with age is still clearly observed. The trend with age seen in Fig.~\ref{age} is consistent with the decline in the disk frequency with age, in particular, the sharp drop observed between $\sim$3 and 5 Myr (e.g., Haisch et al. 2001). This is expected since magnetospheric accretion requires a close-in disk located within the corotation radius (e.g., Hartmann 2005). There can be passive disks which do not show any accretion activity, but not accretors which do not possess disks. The data plotted in Fig.~\ref{age} extends only upto 5 Myr. The oldest accreting BD known to date is the $\sim$8--10 Myr old object 2MASS J1207334-393254 (2M1207), located in the TW Hydrae association (Gizis 2002). This object shows a broad, asymmetric H$\alpha$ profile with a red-shifted absorption component which is likely due to the presence of a highly inclined disk around it (e.g., Mohanty et al. 2005). New accretion activity results for 2M1207 based on X-Shooter spectroscopy are presented in the Stelzer et al. contribution in this proceeding.

\section{Outflow Activity in Brown Dwarfs}

During the active accretion phase, CTTSs are accompanied by energetic winds and outflows, the most prominent diagnostics of which are broad, blue-shifted forbidden emission lines (FELs), and blue-shifted absorption features in both the H$\alpha$ and Na D profiles (e.g., Calvet et al. 1992; Hartigan et al. 1995). The correlation observed between forbidden line emission and disk accretion in CTTSs suggests that the winds and outflows are driven by accretion (e.g., Edwards et al. 1994). The presence of outflow activity in BDs was first suggested by strong FELs observed in the spectrum of the actively accreting late-type (M6.5--M7; 35--72 $M_{Jup}$) object LS-RCrA 1, located in the R Coronae Australis star-forming region (Fern\'{a}ndez \& Comer\'{o}n 2001). However, no apparent outflow extensions could be resolved in the spectrum. It is challenging to directly resolve a BD outflow, since the FEL critical density region is estimated to be on very small spatial scales of $\sim$100 mas, and is located 3--10 times closer to the central source compared to CTTSs (e.g., Whelan et al. 2005). Recent application of the spectro-astrometry technique to high-resolution spectra for BDs has made it possible to recover the spatial offsets in the FEL region. In this technique, high-resolution spectra are obtained at two different slit positions of North-South (P.A. = 0$\degr$) and East-West (P.A. = 90$\degr$) orientations. A positive spectro-astrometric offset in a position-velocity diagram corresponds to extended emission in a northerly/easterly direction, while a negative spatial offset corresponds to a southerly/westerly direction, for the N-S and E-W slit positions, respectively (e.g., Whelan et al. 2005). 

\begin{figure}
\resizebox{\hsize}{!}{\includegraphics{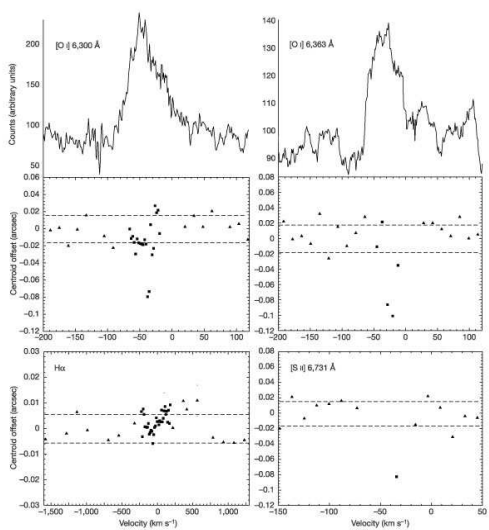}}
\caption{
\footnotesize Line profiles and spectro-astrometric plots for the [OI]$\lambda\lambda$6300,6363 doublet (top and middle panels) and H$\alpha$ and [SII]$\lambda$6731 lines (bottom panel) for $\rho$-Oph 102. Source: Whelan et al. (2005). 
}
\label{oph102}
\end{figure}

\subsection{$\rho$-Oph 102}

\begin{figure}
\resizebox{\hsize}{!}{\includegraphics{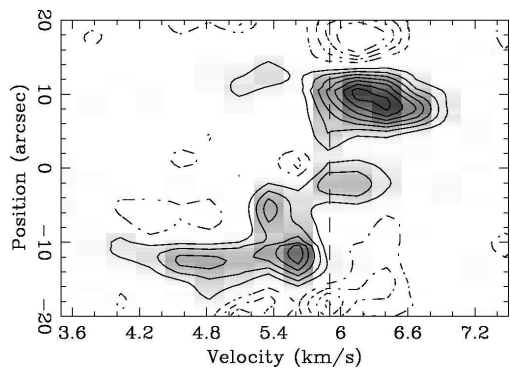}}
\caption{
\footnotesize The CO $J$=2-1 position-velocity diagram for $\rho$-Oph 102. Both blue- and red-shifted components show a range in velocity in their structure. Source: Phan-Bao et al. (2008). 
}
\label{phan}
\end{figure}

The first BD outflow confirmed using the spectro-astrometry technique was for the $\sim$60$M_{Jup}$ object $\rho$-Oph 102, located in the $\rho$-Ophiuchi cloud (Whelan et al. 2005). The centroids of all measurable FELs for this BD show negative offsets with respect to the continuum, and reach a maximum at a blue-shifted radial velocity of $\sim$-40 km s$^{-1}$ (Fig.~\ref{oph102}). The absence of a red-shifted outflow component suggested the presence of an obscuring disk, as noted earlier in CTTS outflow sources (e.g., Hirth et al. 1997). No clear spatial offsets were observed in the H$\alpha$ profile, indicating that most of the H$\alpha$ emission arises from accretion activity. Later, Phan-Bao et al. (2008) reported the observation of a CO $J$=2-1 bipolar molecular outflow driven by $\rho$-Oph 102. Two spatially resolved blue- and red-shifted CO components were observed symmetrically displaced on opposite sides of the BD (Fig.~\ref{phan}). The outflow has a small spatial scale, of about 20$\arcsec$. Based on radiative transfer modeling of the optical to millimeter spectral energy distribution, Phan-Bao et al. (2008) estimated a disk inclination angle of 63$\degr$-66$\degr$ and a projected disk radius of 32-36 AU, and it was reasoned that since the projected disk radius is larger than the jet length, the red-shifted component of the jet is obstructed by the disk in our line of sight. A molecular outflow is indicative of the presence of an underlying jet (e.g., Downes \& Cabrit 2007). This was thus a notable discovery for a BD outflow. 

\begin{figure}
\resizebox{\hsize}{!}{\includegraphics{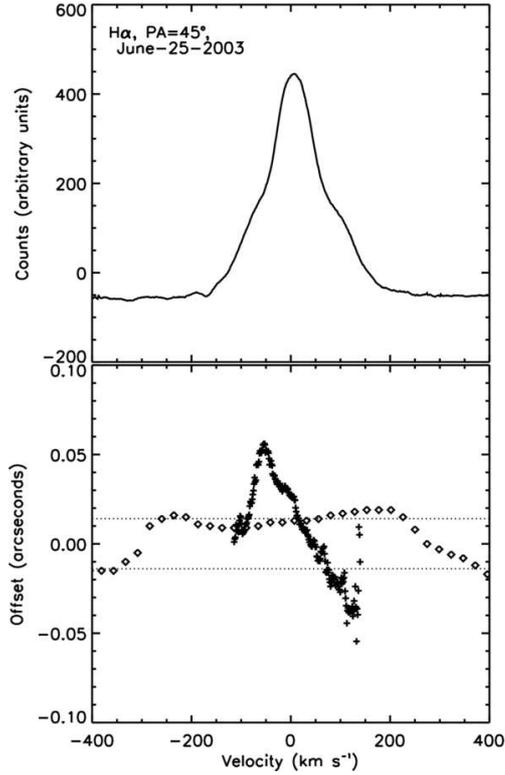}}
\caption{
\footnotesize Spectro-astrometry of the H$\alpha$ line for LS-RCrA 1. Offsets in the wings can be seen at velocities of -50 and +100 km s$^{-1}$. Source: Whelan et al. (2009a). 
}
\label{lscra}
\end{figure}

\subsection{Outflow signatures in H$\alpha$}

A unique case in the field of BD outflows is the object LS-RCrA 1, for which outflow signatures have been observed in the H$\alpha$ profile (Whelan et al. 2009a). This object is an intense accretor, with an H$\alpha$ 10\% width of $\sim$300 km s$^{-1}$ (Fern\'{a}ndez \& Comer\'{o}n 2001). Blue- and red-shifted ``humps'' are observed in the H$\alpha$ profile, which are revealed as spatial offsets at radial velocities of $\sim$-50 km s$^{-1}$ and +100 km s$^{-1}$ (Fig.~\ref{lscra}). Interestingly, spectro-astrometric analysis of the FELs showed blue-shifted spatial offsets only, indicating that the FELs trace only the blue component while both the blue and red outflow lobes are traced by H$\alpha$. This was explained by obscuration of the red-shifted outflow component by the accretion disk, while the red-shifted lobe in H$\alpha$ is likely visible through a $\sim$7 AU sized inner hole in the disk (Whelan et al. 2009a). Furthermore, there was evidence for the presence of both a low-velocity component (LVC) and high-velocity component (HVC) in the outflow, such that the [OI]$\lambda$6300 line at a radial velocity of -5 km s$^{-1}$ mainly traces the LVC, the [NII] line traces the HVC only (centroid at -22 km s$^{-1}$ with the wing extended to -75 km s$^{-1}$), while the [SII]$\lambda$6731 line with a line peak at -13 km s$^{-1}$ and wings extended to -40 km s$^{-1}$ traces both the LVC and HVC. The main highlight of the spectro-astrometric analysis for LS-RCrA 1 was that the H$\alpha$ profile can also trace outflows driven by BDs. 

\begin{figure}
\resizebox{\hsize}{!}{\includegraphics{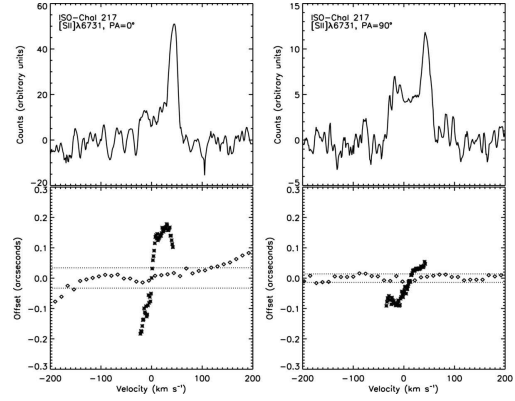}}
\caption{
\footnotesize Spectro-astrometric analysis of the [SII]$\lambda$6731 line for ISO-ChaI 217, with asymmetries observed between the blue- and red-shifted lobes. Source: Whelan et al. (2009b). 
}
\label{iso217}
\end{figure}

\subsection{Asymmetrical outflow}

Another interesting BD outflow case is ISO-ChaI 217, which shows asymmetries in the outflow lobes (Whelan et al. 2009b; Joergens et al. 2012). At an estimated mass of $\sim$80 $M_{Jup}$ (Muzerolle et al. 2005), this object lies at the VLMS/BD boundary. Spectro-astrometric analysis shows the presence of a bipolar outflow, with both blue- and red-shifted components detected in the [SII]$\lambda$$\lambda$6716,6731 and the [OI]$\lambda$$\lambda$6300,6363 FELs at an average radial velocity of -16 and +32 km s$^{-1}$ (Whelan et al. 2009b). There are prominent asymmetries in the blue- and red-shifted lobes traced by the [SII]$\lambda$$\lambda$6716,6731 lines, such that the red-shifted lobe is brighter, has higher electron density, and is faster (a factor of $\sim$2 difference in radial velocity) compared to the blue-shifted lobe (Fig.~\ref{iso217}). Asymmetrical outflow lobes have been observed in CTTS such as RW Aur, for which the redder lobe is brighter and slower than the blue lobe (Hirth et al. 1994). Such similarities suggest similar outflow mechanisms in the sub-stellar mass regime as observed for higher mass stars. 

\subsection{Lowest mass outflow source}

\begin{figure}
\resizebox{\hsize}{!}{\includegraphics{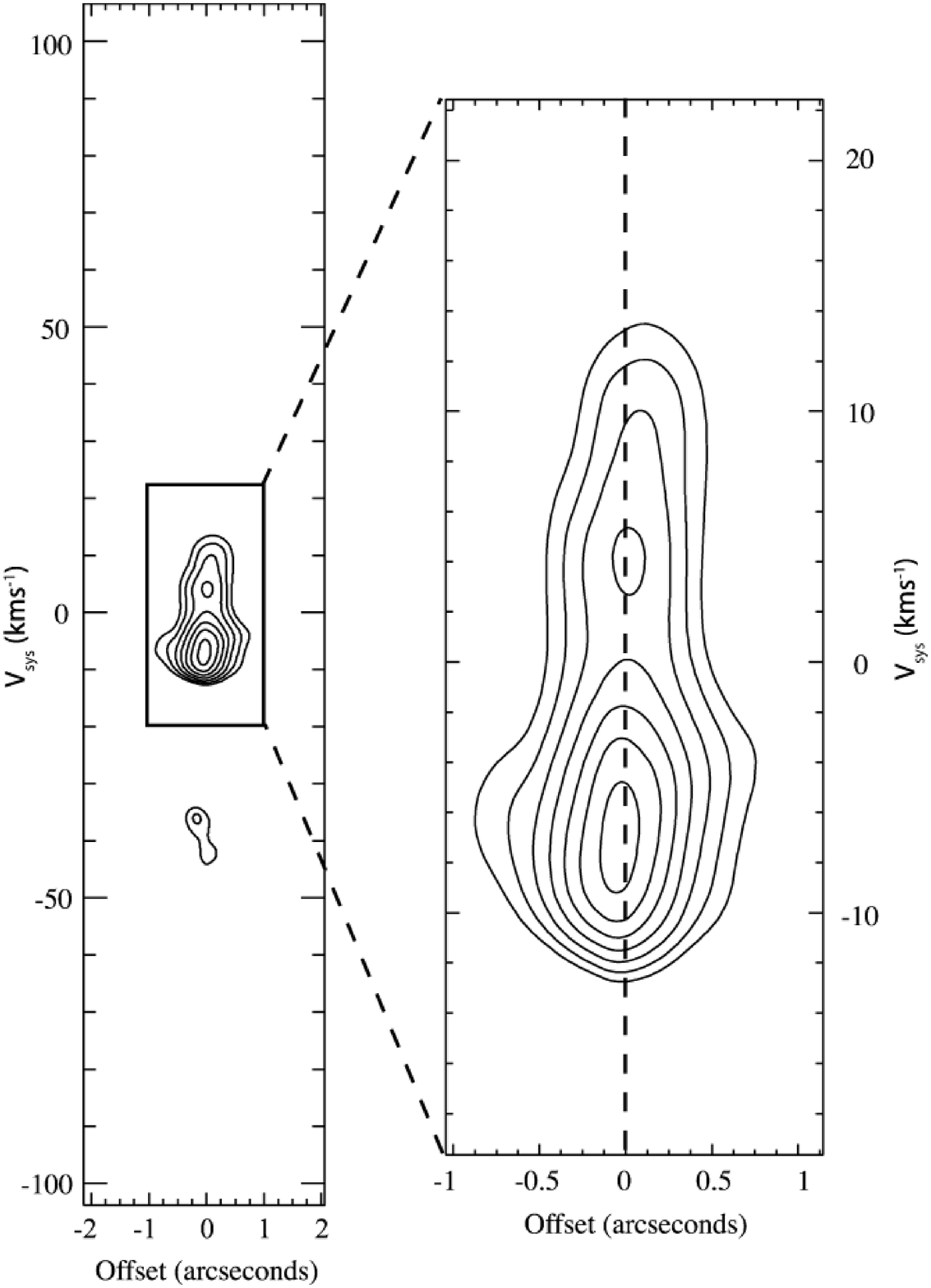}}
\caption{
\footnotesize The position-velocity diagram of the [OI]$\lambda$6300 line for 2M1207, showing blue- and red-shifted offsets at velocities of $\sim$-8 and +4 km s$^{-1}$, respectively. Source: Whelan et al. (2007). 
}
\label{1207}
\end{figure}

The spectro-astrometry technique has been particularly valuable in resolving spatially extended FEL regions in relatively older BDs such as 2M1207. This $\sim$8--10 Myr old $\sim$24 $M_{Jup}$ mass object is an active accretor (e.g., Mohanty et al. 2005), and the lowest mass outflow source known to date. A spectro-astrometric analysis of the FEL region for this BD revealed a faint bipolar outflow, with a blue-shifted component to the [OI]$\lambda$6300 FEL seen at a radial velocity of -8 km s$^{-1}$, and a red-shifted component at +4 km s$^{-1}$ (Fig.~\ref{1207}; Whelan et al. 2007). The discovery of a spatially resolved bipolar outflow for 2M1207 highlights the robustness of the spectro-astrometry method, as well as of the presence of outflow activity down to masses as low as $\sim$0.02 $M_{\odot}$. 

\subsection{Accretion versus outflow activity}

To date, there are very few cases of spatially resolved outflows observed for BDs ($\sim$8--10 in total). The mass outflow rate measured for these BDs range between $\sim$1$\times$10$^{-9}$--6$\times$10$^{-10}$ $M_{\odot}$ yr$^{-1}$ (Whelan et al. 2009b). However, the degeneracies due to the method used for measuring the mass loss rate, the estimates for the critical and electron densities, the jet velocity, etc., could result in an order of magnitude difference in the outflow rate. A prime example is LS-RCrA 1, for which the outflow rate measured from the total density in the outflow (method A described in Whelan et al. 2009b) is 2.4$\times$10$^{-9}$ $M_{\odot}$ yr$^{-1}$, compared to 2--6$\times$10$^{-10}$ $M_{\odot}$ yr$^{-1}$ measured from the [OI]$\lambda$6300 and [SII]$\lambda$6731 line luminosities. Similar spread of an order of magnitude is found for the mass accretion rate measurements (e.g., Rigliaco et al. 2012). Overall, the mass outflow rates for BDs are comparable to the mass accretion rates, however, the ratio $\dot{M}_{out}$/$\dot{M}_{acc}$ could be larger than the 1--10\% estimated for CTTSs (e.g., Hartigan et al. 1995). As noted in Whelan et al. (2009b), there may be a potential observational bias in terms of selecting only those BDs for a spectro-astrometric analysis which show strong emission in the [OI]$\lambda$6300 FEL, thus selecting BD outflows with a high mass loss rate of 10$^{-9}$--10$^{-10}$ $M_{\odot}$ yr$^{-1}$. Among CTTSs, a large scatter is observed about the linear fit in the $\dot{M}_{out}$ versus $\dot{M}_{acc}$ correlation (e.g., Hartigan et al. 1995). Future work with more sensitive instruments such as X-Shooter will be important in observing an unbiased sample and discovering outflows for weaker BD accretors, with accretion rates of $\leq$10$^{-11}$ $M_{\odot}$ yr$^{-1}$, thereby confirming that jets and outflows can also be driven at very low mass accretion rates. 









\begin{acknowledgements}

I would like to thank the organizers for providing a fun-filled learning experience at the conference. 

\end{acknowledgements}

\bibliographystyle{aa}

\end{document}